# High-throughput screening of encapsulated islets using wide-field lens-free on-chip imaging

Yibo Zhang,[1,2,3] Michael Alexander,[4] Sam Yang,[1] Yinxu Bian,[1] Elliot Botvinick,[5,6] Jonathan R.T. Lakey,[4,5] Aydogan Ozcan[1,2,3,7*]

[1]*Electrical & Computer Engineering Department, University of California, Los Angeles, CA, 90095, USA.*
[2]*Bioengineering Department, University of California, Los Angeles, CA, 90095, USA.*
[3]*California NanoSystems Institute (CNSI), University of California, Los Angeles, CA, 90095, USA.*
[4]*Department of Surgery, University of California Irvine, CA 92868, USA.*
[5]*Department of Biomedical Engineering, University of California Irvine, CA 92617, USA.*
[6]*The Edwards Lifesciences Center, University of California Irvine CA 92617, USA.*
[7]*Department of Surgery, David Geffen School of Medicine, University of California, Los Angeles, CA, 90095, USA.*
*\*Corresponding author: ozcan@ucla.edu*

http://innovate.ee.ucla.edu/   http://org.ee.ucla.edu/

**Islet microencapsulation is a promising solution to diabetes treatment, but its quality control based on manual microscopic inspection is extremely low-throughput, highly variable and laborious. This study presents a high-throughput islet-encapsulation quality screening system based on lens-free on-chip imaging with a wide field-of-view of 18.15 cm², which is more than 100 times larger than that of a lens-based optical microscope, enabling it to image and analyze ~8,000 microcapsules in a single frame. Custom-written image reconstruction and processing software provides the user with clinically important information, such as microcapsule count, size, intactness, and information on whether each capsule contains an islet. This high-throughput and cost-effective platform can be useful for researchers to develop better encapsulation protocols as well as perform quality control prior to transplantation.**

Type 1 diabetes (T1D) originates from an autoimmune disorder that destroys the insulin-secreting β-cells in the islets of Langerhans in the pancreas. It affects more than 420 million people worldwide, as estimated in 2014 [1], which continues to increase. Islet transplantation has been employed to treat T1D using pancreas transplantation from deceased donors; however, this is limited owing to the rejection by the recipient's immune response even when chronic immunosuppression is employed [2,3]. The lack of suitable human donor organs also leads to delays in treatment [4–6]. In order to overcome some of these issues, islet microencapsulation has been developed [7–10], which uses a biocompatible semipermeable polymer to encase the islets, forming microcapsules with a typical diameter in the range of a few hundred micrometers to over one millimeter [11]. The semipermeable membrane blocks the immune cells from attacking the implanted islets, while allowing nutrients and insulin to pass through. Islet encapsulation holds potential to replace pharmaceutical immune suppression, which is the standard of care for the recipient of an islet transplant. Thus, islet encapsulation may not only improve outcomes following transplantation, but perhaps may also allow for the use of xenografts, such as islets harvested from pigs, thereby resolving the unavailability problem of donor organs [12,13].

The morphology of the microcapsules as well as the size and position of the islets within the microcapsules are all important to the success of the treatment [11,14–16]. Therefore, a rigorous quality control through a microscopic inspection of the microencapsulation quality is an integral step before implantation into the patient. However, as hundreds of thousands or even millions of encapsulated islets are needed by each patient [17], thousands of encapsulated islets should be sampled to obtain accurate statistics about their morphological variations and overall quality. Nevertheless, the current inspection approach using conventional optical microscopy is limited by its small field-of-view (FOV) [18], which provides challenges to image large numbers of encapsulated islets in a high-throughput manner, posing a practical challenge to efficient quality control. Owing to the relatively large size of each microcapsule (a few hundred micrometers to over one millimeter), a high-throughput imaging system is desired with a wide FOV that is at least on the order of several square centimeters with a resolution that is sufficient to resolve the fine morphology of the microcapsules and islets.

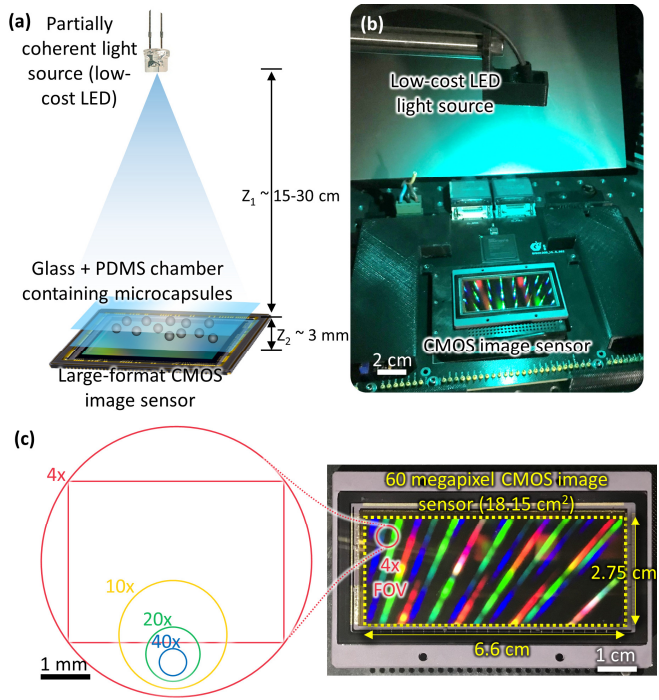

Fig. 1. Ultra-wide-FOV microscopic imaging platform, based on lens-free on-chip holographic microscopy for high-throughput analysis of encapsulated islets. (a) Schematic of the optical setup. (b) Photograph of the experimental optical setup. (c) FOV comparison between typical microscope objective lenses and lens-free on-chip microscopy. The lens-free microscope's FOV (~18.15 cm$^2$) is ~76 times larger than that of a 4× microscope (red circle in the left) when observing from the eyepiece, and ~159 times larger than the FOV of the same 4× microscope when using a digital microscope camera (with an aspect ratio of 4:3, red rectangle in the left part).

In this work, we present an ultra-high-throughput lens-free on-chip imaging platform for rapid screening of alginate encapsulated islets over an FOV of ~18.15 cm$^2$, i.e., two orders of magnitude larger than that of a typical 4× microscope objective lens, enabling to image and analyze over 8,000 microcapsules in a single frame over an image acquisition time less than 1 s (Figure 1). This lens-free on-chip microscopy technique [19–40] is based on the principle of digital in-line holography with a unit fringe magnification geometry, which decouples the imaging FOV and resolution. Therefore, it can simultaneously achieve an extremely large FOV with a high resolution. Moreover, the lens-free on-chip microscopy eliminates the use of lenses in the imaging setup; it uses holography to reconstruct the image. Consequently, the optical setup is very simple and cost-effective to build [21,25,33–35]. Figures 1(a) and 1(b) show the optical setup that we specifically built for screening of encapsulated islets with an ultra-wide FOV, using a large-format 60-megapixel complementary-metal-oxide-semiconductor (CMOS) image sensor (pixel size: 5.5 µm) for hologram acquisition and a 505-nm light-emitting diode (LED) for illumination. In order to provide a uniform illumination over such a wide FOV with a relatively small light-source-to-sample distance ($z_1$), the LED's plastic lens is sanded and polished into a flat profile close to the active region of the LED to significantly increase its emission angle (see Figure 1 (a-b)), leading to a minimum $z_1$ distance of ~15 cm. A custom-designed sample chamber made from glass and polydimethylsiloxane (PDMS) is used to hold the microcapsules, placed directly on top of the CMOS image sensor, leading to a sample-to-image-sensor distance ($z_2$) of ~3 mm. Following the in-line hologram acquisition, a computational framework is used to identify, classify, and characterize each encapsulated islet within the wide-field image. In addition to providing a high-throughput tool to inspect the quality of microcapsules, this platform offers the possibility to preselect high-quality encapsulated islets before implantation, when combined with micromanipulation techniques.

The large numbers (e.g., ~8,000) of microcapsules imaged in each frame make manual analysis labor intensive, prone to error and ultimately unfeasible. Therefore, following the image acquisition, the acquired wide-field hologram is rapidly analyzed using a custom-developed MATLAB-based automated classification and sizing algorithm, which measures clinically relevant characteristics of individual microcapsules and provides their statistics. We demonstrate the capability of our platform for sizing, as the microcapsules' size is an important parameter that affects their biocompatibility, response to glucose, capability to secret insulin, and capability to withstand osmotic and physical stresses [11,14–16]. In addition, we demonstrate the capability of this platform for the identification of intact microcapsules with islets, intact microcapsules without islets, and microcapsules with islet protrusion, which is essential to ensure a successful transplantation [17]. In addition to these, various other morphological parameters can be obtained by introducing additional image processing steps.

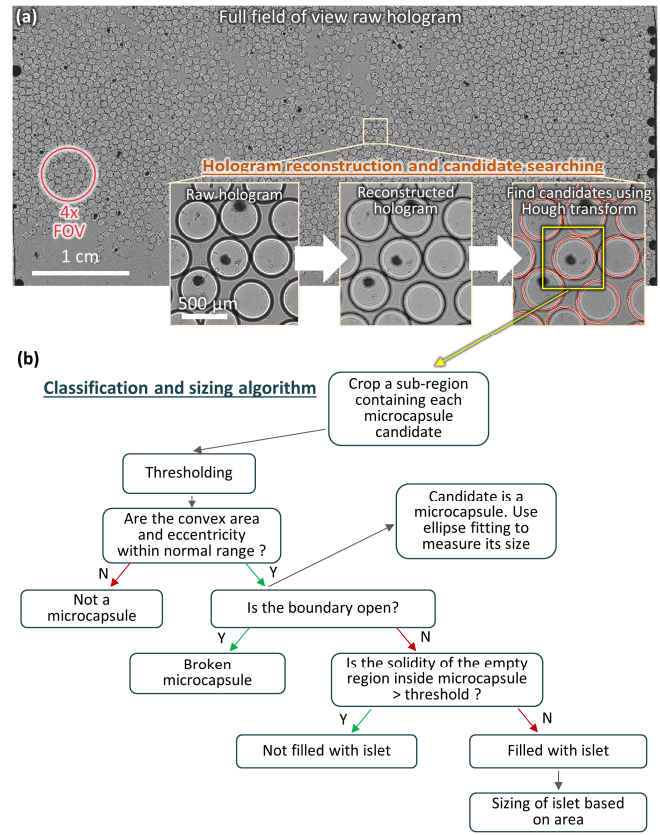

Fig. 2. A custom-written algorithm is used to detect microcapsules, measure their sizes and sort them into different classes for clinical use. (a) A full-FOV hologram (FOV = 18.15 cm$^2$) is back-propagated to the optimal object plane corresponding to the microcapsules, and microcapsule candidates are identified using Hough transform. (b) Each of the microcapsule candidates is individually analyzed using a classification and sizing algorithm.

As part of our processing, the hologram is first back-propagated to the object plane corresponding to the microcapsules [30,41]. Hough transform [42] is then used to detect the round or elliptical objects within the back-propagated hologram amplitude, which are used as the initial "candidates" for microcapsules. Based on the coordinates and diameters of these detected candidates, a square window, 10% larger than the diameter, is used to crop out the image of each individual candidate. We then apply a threshold and find the connected region with the largest convex area, i.e., area of the convex hull, which should correspond to the boundary of the microcapsule. If the connected region's convex area and eccentricity are outside an extended normal range (i.e., the convex area is smaller than the area of a circle with a 200 μm diameter or larger than the area of a circle with a 1500 μm diameter, or the eccentricity is larger than 0.8), they are considered false detections and are digitally filtered out, while those within this range are considered true microcapsules, and are used for sizing and classification. In the sizing stage, the algorithm tracks the boundary of each microcapsule, yielding a set of points ($x$–$y$ coordinates) along the boundary. It then feeds this set of $x$–$y$ coordinates to an ellipse fitting algorithm [43], outputting its major axis, minor axis, and eccentricity. In the classification stage, based on whether the microcapsule's boundary is open or not, first, the algorithm classifies the microcapsule as intact/broken. For intact microcapsules, if the empty region within the microcapsule is larger than a preset threshold that was manually tuned, the algorithm classifies it as empty; otherwise, the algorithm classifies it as islet-containing. For islet-containing microcapsules, the islets' sizes are also measured based on their areas.

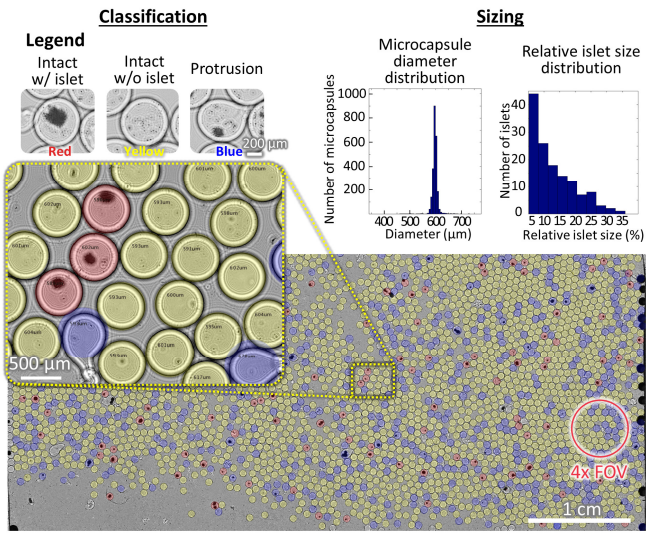

Fig. 3. Results obtained using the microcapsule classification and sizing algorithm. Classification: the three different classes of microcapsules (i.e., intact with islet, intact without islet, and protrusion) are color-coded in the final image for better visualization. Legend (top left): red: intact with islet, yellow: intact without islet, and blue: protrusion. Sizing: the diameter (the average of the major and minor axes) of each microcapsule is shown (see the magnified image). The algorithm can also output statistics about various microcapsule parameters, such as the diameter and relative islet size (islet area over each microcapsule area) distributions, shown in the top right part. The black circular structures that appear at the right edge of the FOV are air bubbles trapped within the channel, which are digitally rejected. A wider microchannel can eliminate the appearance of such bubbles within the lens-free FOV.

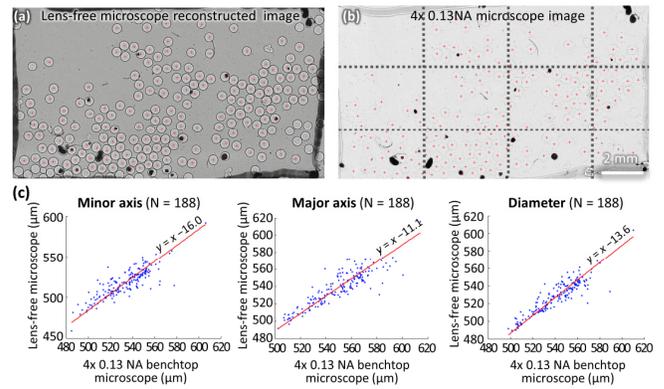

Fig. 4. Sizing accuracy of the lens-free microscope. (a) A zoomed-in lens-free microscope image and (b) a digitally-stitched 4× 0.13 NA microscope objective image of the same region of interest. For comparison, the lens-free microscope image is cropped to a pre-defined smaller region, and multiple 4× microscope images are stitched together to match the corresponding FOV. The dashed grids in (b) represent the individual FOVs of the 4× microscope objective. The red crosses represent corresponding microcapsules within the two images. (c) The comparison of the minor axis, major axis, and diameter (the average of the minor and major axes) obtained from the lens-free microscope and a 4× microscope objective shows a good agreement.

The above described microcapsule classification and sizing algorithm was used to obtain various morphological parameters of individual microcapsules, as well as their statistics, as shown in Figure 3. In the displayed reconstructed lens-free amplitude image, we used different colors to overlay on the grayscale image, in order to represent the different classes of microcapsules and facilitate their visualization. The classification algorithm could accurately identify microcapsules with different traits. The diameter (the average of the major and minor axes) of each microcapsule is shown, and its statistical distribution is plotted as a histogram. The statistical distribution of the relative islet size (i.e., islet area over each microcapsule area) is also shown in Figure 3. Other parameters, such as the relative position of the islets with respect to the microcapsule, can also be analyzed by introducing additional data processing steps.

As the encapsulated islets are three-dimensional objects, their reconstructed holographic images contain information from out-of-focus parts of the microcapsules and islets, making the boundaries of the microcapsules slightly thicker, as shown in Figure 2 (a) (reconstructed hologram). The low coherence of the LED illumination also leads to a degradation in resolution, which can be improved using a light source with a higher coherence, such as a spatially and spectrally filtered LED or a laser diode [33].

Next, we performed an experiment to compare the sizing results obtained with the lens-free microscope and a 4× 0.13 NA

microscope objective. A zoomed-in region that contains 188 microcapsules is defined on the sample chamber, which is imaged by both imaging modalities (Figs. 3 (a) and (b)). The microcapsules in the lens-free reconstruction and 4× objective images are manually matched to undo the effect of sample displacement during the transportation between the two microscopes. As shown in Figure 4 (c), the average differences between the values of the major axis, minor axis, and diameter measured by the lens-free microscope and 4× benchtop microscope are 16.0 µm, 11.1 µm, and 13.6 µm, respectively, i.e., smaller than ~3% of the average diameter of the microcapsules in the sample, providing a good agreement between the results obtained using the lens-free microscope and a benchtop microscope.

In conclusion, we presented an ultra-wide-FOV lens-free on-chip microscopy platform for a high-throughput analysis of encapsulated islets as a quality control tool. A low-cost LED and a large-format 60-megapixel CMOS image sensor were used to construct a cost-effective setup, with a FOV of 18.15 cm$^2$, which is two orders of magnitude larger than that of a 4× objective lens. Over 8,000 microcapsules can be analyzed using the proposed platform in a single shot, with an image acquisition time that is less than 1 s. In addition, we developed an automated high-throughput microcapsule classification and sizing algorithm to be used in conjunction with the imaging hardware, which detects the microcapsules within the FOV, classifies each microcapsule into clinically-relevant classes, and measures their sizes. Furthermore, the algorithm outputs the statistical distribution of these parameters within the sample population, which can provide important information regarding the quality of the analyzed microcapsule batch. The unique capability of the presented platform to rapidly and automatically analyze an extremely large number of microcapsules, in addition to its low cost and simple design, reveals its potentials as a powerful tool for pre-transplant quality control as well as for the improvement of the encapsulation procedures. Combined with high-throughput micromanipulation techniques, this platform could be used for pre-selection of high-quality encapsulated islets to improve treatment outcomes.

**Funding.** The authors acknowledge the support of NSF Engineering Research Center (ERC, PATHS-UP), the Howard Hughes Medical Institute (HHMI), UC Irvine Department of Surgery, Juvenile Diabetes Research Foundation (JDRF).